\documentclass[twocolumn,showpacs,floatfix,prb]{revtex4}
\usepackage{graphicx}
\usepackage{dcolumn}
\usepackage{bm}
\usepackage[latin1]{inputenc}
\usepackage{float}
\usepackage{soul}
\usepackage{dsfont}
\usepackage{amsmath}

\newcommand{\bfa}{{\cal W}\!\!\!\!\!\!{\cal W}}

\RequirePackage{color}

\def\ket#1{|\!#1\rangle}
\def\bra#1{\langle#1\!|}

\begin{document}

\title{Spin-polarization of entangled and mixed electron states in a beam splitter geometry coupled to an electron reservoir}

\author{Luis A. Gonz\'alez-\'Arraga}
\affiliation{Laboratorio de F\'isica Estad\'istica de Sistemas Desordenados, Centro de F\'isica, Instituto Venezolano de Investigaciones
Cient\'ificas. IVIC, Apartado 21827,Caracas 1020 A, Venezuela}

\author{Bertrand Berche}
\affiliation{Laboratorio de F\'isica Estad\'istica de Sistemas Desordenados, Centro de F\'isica, Instituto Venezolano de Investigaciones Cient\'ificas.
IVIC, Apartado 21827,Caracas 1020 A, Venezuela}
\affiliation{Statistical Physic{s} Group, P2M, Institut Jean Lamour, Nancy Universit\'e, BP70239, F-54506 Vand\oe uvre les Nancy, France}

\author{Ernesto Medina}
\affiliation{Laboratorio de F\'isica Estad\'istica de Sistemas Desordenados, Centro de F\'isica, Instituto Venezolano de Investigaciones Cient\'ificas.
IVIC, Apartado 21827,Caracas 1020 A, Venezuela}
\affiliation{Statistical Physic{s} Group, P2M, Institut Jean Lamour, Nancy Universit\'e, BP70239, F-54506 Vand\oe uvre les Nancy, France}

\date{\today}

\begin{abstract}
We study the spin polarization of mixed and entangled electron states in a four probe/beam splitter geometry with local
Rashba and Dresselhaus interactions. A pair of maximally entangled electrons collides with the beam splitter and enters into two perpendicular branches of length $L$, composed of spin-orbit active materials (gate confined 2D electron gas). One of the branches is connected to an electron reservoir that acts as a source of decoherence by either behaving as a voltage probe or as a controlled source or sink of current at fixed voltage. Such decoherence source is used to modify the entropy of an unpolarized incoming state in order to generate electron polarization at one or both output branches. The degree of entanglement of the global state and the spin polarization is computed for the outgoing electrons as a function of the coupling to the electron reservoir. Experimentally available spin-orbit strength at the beam splitter arms, for arm lengths of a few micrometers, is able to modulate spin polarization up to 80\% in particular spin axes. The Dresselhaus and Rashba coefficients play a symmetric role in modulating the polarization. Significantly less polarization is achieved for incoming mixed states due to the local operation of the reservoir.
\end{abstract}
\maketitle

\section{Introduction}

One of the most coveted resources in the new field of spintronics is the availability of a controlled source of spin polarized electrons
\cite{DasSarma}. One main focus of application is spin injection, frequently from a ferromagnetic contact into a semiconductor where spin manipulation circuits or memory applications are implemented\cite{Hammar,Gardelis}. The main drawback of such approach is the ferro-semiconductor mismatch that reduces spin injection efficiencies. Another focus of interest are spin-centered quantum information applications that require precise manipulation of spin states\cite{DiVincenzo,DasSarma2}.

Spin control through electrical means, contemplating spin-orbit active media, is desirable since gate control is a common tool in large scale integration. Recently, spin coupled electron interference has been considered to generate spin polarization from unpolarized
sources\cite{Nitta,Ionicioiu,Ting,Molnar,Hatano,Chen,Lopez}, developed from many studies of spin-orbit active media in multiply connected geometries\cite{Oreg,YiQianSu,Nikolic,Frustaglia,Zulicke,Foldi}. Such a concept makes spin-polarized electrons available within the semiconducting medium avoiding interface problems and offering refined control through a number of experimentally accessible parameters, such as gate voltages, external magnetic fields, material strains and geometrical configurations.

Interferometric proposals for spin polarization, such as those considered above, depend on phase coherence, so their performance will be compromised if one considers finite temperatures and connections with a reservoir\cite{Santos}. Size reduction of device filters, so that they are within a phase coherence length makes operation only possible at temperatures in the mK range for systems of micrometers in length.
Even satisfying requirements of size and temperature, phase coherence can be compromised by orbital coupling to the spin degree of freedom when more than one sub-band is available\cite{Nikolic2} in quasi-one dimensional leads.  

In this work we propose to take advantage of limited coherence by exploring how decoherence can spin polarize electrons through entropy changes of the incoming state due to reservoir coupling. We use this concept in order to polarize initially unpolarized entangled and mixed states. The degradation of entanglement of a Bell pair, and the polarization of initially unpolarized mixed electron states, is achieved by connecting a spin-orbit active region to an electron reservoir. As a source of decoherence, we use the reservoir model of
B\"uttiker\cite{Buttiker,Buttikeretal}. Such a local model, introduces the action of a reservoir using the scattering matrix description, contemplating temperature, the Fermi level or a voltage, with a controlled coupling to the system.  A symmetry based argument to understand the manipulation of the global polarization with a reservoir, is that it breaks time reversal symmetry in the system, thus allowing for changes in the global polarization. If this were not the case, the proposed setup could only redistribute the polarization of the input state.

Changes in the state entropy of the system coupled to the reservoir offers a clear view as to the actual operation of Buttiker's model that is interesting in itself. The treatment being completely coherent, only introduces complex phase relations between the lateral lead and the system through continuous energy integration. The zero current condition in the connecting lead and the fact that we couple the system with a reservoir exhibiting a continuous spectrum leaves the state entropies unchanged. Nevertheless, the presence of two spin channels leaves the zero current condition adjustable, so real decoherence can be achieved by a dissipative or lossy non-unitary coupling. Decoherence implemented this way, will change the entropy of interferometer state offering a tool for generating polarized electrons.

Feve \emph{et al}\cite{Feve} have established the influence of spin-orbit coupling effects in two dimensional systems within the framework of the Landauer-B\"uttiker coherent scattering formalism, defining the creation and annihilation operators for the Rashba Hamiltonian. They calculated the scattering matrix of a four-port beam splitter, that mixes states with different spin-orbit coupling labels, and determined the dependence on the angle between the port arms. This description is particularly suited for our case once generalized to include 2D Dresselhauss interactions.

The paper is structured as follows. In Section 2, we diagonalize the Hamiltonian with Rashba and Dresselhaus spin-orbit interactions for a 2DEG and obtain the creation operators for confined electrons propagating through this medium following the second-quantized approach of  ref. [\onlinecite{Feve}]. In Section 3, we study the process of scattering of a pair of maximally entangled electrons by a beam splitter device followed by two perpendicular arms of spin-orbit active media coupled to an electron reservoir. We compute the change in the concurrence as a function of the reservoir coupling, decoherence current and temperature sensitivity. In section 4 we analyze the polarization of the outgoing electrons in the entanglement correlated arms of the beam splitter. The controlled sink current and its relation to the system-reservoir coupling is established, and shown to generate strong spin polarization. In Section 5, we study the process of scattering of electrons that are initially unpolarized in a completely mixed state, by the same device. We conclude with a summary of our findings.

\section{Rashba-Dresselhaus Hamiltonian. Creation and annihilation operators}
The Hamiltonian for an electron in a Rashba-Dresselhaus spin-orbit active material is given by
\begin{equation}
H=\frac{\hbar^2 k^2}{2m}-\alpha \hbar (k_{x}\sigma^{y}-k_{y}\sigma^{x})-\beta \hbar (k_{y}\sigma^{y}-k_{x}\sigma^{x}),
\end{equation}
where $\alpha$ and $\beta$ are the Rashba and Dresselhaus spin-orbit parameters, and $\sigma^i$ is the i-th Pauli matrix. The eigenstates and eigenvalues are
\begin{eqnarray}
\vert \Psi, E_{\pm}\rangle = \frac{1}{\sqrt{2}}\left[\vert \uparrow \rangle  \mp \frac{\alpha e^{-i\theta_{\gamma}}-i\beta e^{i\theta_{\gamma}}}{\sqrt{\alpha^2 + \beta^2 +4\alpha\beta \sin \theta_{\gamma} \cos \theta_{\gamma}}}\vert \downarrow \rangle \right]\nonumber \\
\times e^{i(k_{x}x+k_{y}y)},\nonumber \\
\end{eqnarray}
and
\begin{equation}
E_{\pm}=\frac{\hbar^2 k^2}{2m}\pm \hbar k \sqrt{\alpha^2 +\beta^2 +4\alpha\beta\sin \theta_{\gamma} \cos\theta_{\gamma}},
\label{autovaloresSO}
\end{equation}
where $\vert \uparrow \rangle, \vert \downarrow \rangle$ are the eigenstates of $\sigma_z$ with eigenvalues $\pm 1$ and $\theta_{\gamma}$ is the angle between the electron wavevector and here the laboratory $y$ axis.
The dispersion relation given by Eq.\ref{autovaloresSO} allows us to explore the symmetries of the Hamiltonian. Time reversal
changes the sign of the wavevector $k$ and of the spin angular momentum $\sigma$, leaving the coordinates untouched
while spatial inversion changes $k$ and the spatial coordinates but not the spin. The lack of spatial inversion symmetry
with time reversal symmetry preserved results in $E_{\sigma}(k) \neq E_{-\sigma}(k)$.

If the electron propagates in the $y$ direction, $\theta_{\gamma}=0$, with confinement in the $x$ direction, for a single transverse sub-band, then
\begin{equation}
\vert \Psi, E_{\pm}\rangle = \frac{1}{\sqrt{2}} \Phi_{x}(x,y) \left[ \vert \uparrow \rangle  \mp \frac{\alpha -i\beta}{\sqrt{\alpha^2 + \beta^2}} \vert \downarrow \rangle \right]e^{ik_{\pm}y},
\label{confinex}
\end{equation}
where $\Phi_{x}(x,y)$ confines the wavefunction perpendicular to the direction of propagation. The dispersion relation for this configuration is then
\begin{equation}
E_{\pm}(k)=\frac{\hbar^2k_{\pm}^2}{2m_{0}}\pm \hbar k_{\pm} \sqrt{\alpha^2+\beta^2},
\label{EnergyDispersion}
\end{equation}
Fixing the particular energy to $E$, the wavevectors $k_+$ and $k_-$ are given by
\begin{equation}
k_{+}=\frac{m}{\hbar}\sqrt{\alpha^2+\beta^2+\frac{2E}{m}}-\frac{m}{\hbar}\sqrt{\alpha^2+\beta^2},
\end{equation}
\begin{equation}
k_{-}=\frac{m}{\hbar}\sqrt{\alpha^2+\beta^2+\frac{2E}{m}}+\frac{m}{\hbar}\sqrt{\alpha^2+\beta^2}.
\end{equation}
Defining $\kappa\equiv\arctan{\frac{\beta}{\alpha}}$, the creation operators in the spin-orbit basis corresponding to electrons traveling in the $y$ direction {along
leads denoted as 2 and 3 in the following} are written as
\begin{equation}\nonumber
a_{2+}^{\dagger}=a_{3+}^{\dagger}=\frac{\Phi_{x}(x,y)}{\sqrt{2}}\left[a_{3 \uparrow}^{\dagger}-e^{-i\kappa} a_{3\downarrow}^{\dagger}\right]e^{ik_{+}y},
\end{equation}
\begin{equation}\nonumber
a_{2-}^{\dagger}=a_{3-}^{\dagger}=\frac{\Phi_{x}(x,y)}{\sqrt{2}}\left[a_{3 \uparrow}^{\dagger}+e^{-i\kappa} a_{3\downarrow}^{\dagger}\right]e^{ik_{-}y}.
\end{equation}
If on the other hand the electron propagate in the $x$ axis ($\theta_{\gamma}=\frac{\pi}{2}$), confinement operates in the $y$ direction and the wavefunction is
\begin{equation}
\vert \Psi, E_{\pm}\rangle = \frac{1}{\sqrt{2}} \Phi_{y}(x,y) \left[ \vert \uparrow \rangle  \mp \frac{\beta -i\alpha}{\sqrt{\alpha^2 + \beta^2}} \vert \downarrow \rangle \right]e^{ik_{\pm}x}
\label{confiney}
\end{equation}
and the dispersion relation is the same as that of Eq.~\ref{EnergyDispersion}.
The corresponding creation operators in the spin-orbit basis for electrons traveling in the $x$ direction {along leads 1 and 4} are then
\begin{equation}
a_{1+}^{\dagger}=a_{4+}^{\dagger}=\frac{\Phi_{y}(x,y)}{\sqrt{2}}\left[ a_{4\uparrow}^{\dagger}+ie^{i\kappa} a_{4\downarrow}^{\dagger}\right]e^{ik_{+}x},
\end{equation}
\begin{equation}
a_{1-}^{\dagger}=a_{4-}^{\dagger}=\frac{\Phi_{y}(x,y)}{\sqrt{2}}\left[ a_{4\uparrow}^{\dagger}-ie^{i\kappa} a_{4\downarrow}^{\dagger}\right]e^{ik_{-}x},
\end{equation}
The operators for electrons traveling in the negative direction in cables 1 and 4 are
\begin{equation}
b_{1+}^{\dagger}=b_{4+}^{\dagger}=\frac{\Phi_{y}(x,y)}{\sqrt{2}}\left[ b_{4\uparrow}^{\dagger}-ie^{i{\kappa}} b_{4\downarrow}^{\dagger}\right]e^{-ik_{+}x},
\end{equation}
\begin{equation}
b_{1-}^{\dagger}=b_{4-}^{\dagger}=\frac{\Phi_{y}(x,y)}{\sqrt{2}}\left[ b_{4\uparrow}^{\dagger}+ie^{i {\kappa}} b_{4\downarrow}^{\dagger}\right]e^{-ik_{-}x},
\end{equation}
and the operators {$b^{\dagger}_{4+}$ and $b^{\dagger}_{4-}$} are related to $a^{\dagger}_{\gamma \sigma}$ through $b^{\dagger}_{4+}=e^{-i(k_{+}+k_{-})x}a^{\dagger}_{4-}$ and $b^{\dagger}_{4-}=e^{-i(k_{+}+k_{-})x}a^{\dagger}_{4+}$.
\begin{figure}
\includegraphics[width=9cm]{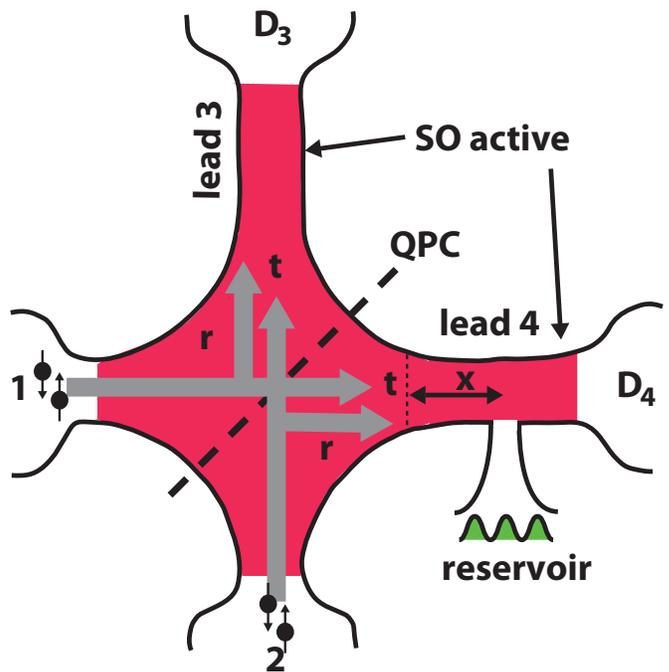}
\caption{\label{fig1} Sketch of the spin-polarizing device. Incoming electrons traveling through leads 1 and 2 are scattered into leads 3 and through the four-port beam splitter QPC implemented within a SO active region of the Rashba and Dresselhaus type. Lead 4 is connected to a voltage probe model electron reservoir. The electrons are collected at detectors $D_{3}$ and $D_{4}$. $x$ denotes the position of the reservoir junction with respect to the QPC.}
\end{figure}
The fermionic operators $a_{\gamma \sigma}^{\dagger}$ and $a_{\gamma \sigma}$ belong to the spin-orbit basis and therefore
create eingenstates of the Rashba-Dresselhaus Hamiltonian. The operators obey the anti-commutation relation
 $\{a_{\gamma \sigma},a_{\gamma^{\prime} \sigma^{\prime}}^{\dagger}\}=\delta_{\gamma \gamma^{\prime}}\delta_{\sigma \sigma^{\prime}}$ .  They are associated with an energy $E_{\sigma}$ according to the Eq.\ref{autovaloresSO}.
It is important to point out that the previous solutions are exact only for single sub-band transmission in the beam splitter arms.
When more than one sub-band is present further considerations must be exercised since the SO interaction couples the sub-bands and the wave functions of Eq.\ref{confinex} and \ref{confiney} are no longer eigenfunctions. Interesting effects arise for more than one sub-band that include additional precessional phases\cite{Egues} and intrinsic decoherence\cite{Nikolic2}. 

The physical conditions for pure treatment of the wave functions to be appropriate are: i) The lead confining potentials are sufficiently shallow to support a single subband ii) If there is a second subband, that the input Fermi energy be smaller than the second transverse mode\cite{Nikolic2} iii) The SO matrix element between subbands times the Rashba coefficient is much smaller than the energy separation between subbands\cite{Egues}. In these cases our approach yields a good description of the transport properties of the leads.

\section{Decoherence of a Bell pair in the beam splitter\label{section3}}
We consider the scattering of a Bell pair with the beam splitter configuration shown in Fig.~\ref{fig1}. The whole configuration is assumed a gate controlled on a 2 dimensional electron gas of spin-orbit active material.  The incoming electrons travel through leads $1$ and $2$ before colliding against the beam splitter, which can be implemented experimentally by a quantum point contact  {(QPC)}\cite{Oliver} (within the SO region). After collision with the beam splitter the electrons enter a region that is made of two perpendicular branches of length $L$ composed of Rashba and Dresselhaus spin-orbit active media (leads $3$ and $4$). Lead $4$ is connected through a lead to an electron reservoir coupled equally to both spinor components in the spin-orbit eigenbasis of the beam splitter arm. Detectors $D_{3}$ and $D_{4}$ collect the outgoing electrons. The state of the incoming electrons is  chosen
{in this section} to be a Bell pair
\begin{equation}
\vert \Psi_{Bell}\rangle =\frac{1}{\sqrt{2}}\left(a_{1+}^{\dagger}a_{2+}^{\dagger}+ a_{1-}^{\dagger}a_{2-}^{\dagger} \right) \vert 0 \rangle,
\end{equation}
written in the spin-orbit basis. The collision with the beam splitter is described by the  $4\times 4 $ scattering matrix:
\begin{equation} S=\left( \begin{array}{cccc}
r\cos  {\vartheta}_{i} & ir\sin  {\vartheta}_{i} & t & 0   \\
ir\sin  {\vartheta}_{i} & r\cos {\vartheta}_{i} & 0 & t\\
t & 0 & r\cos  {\vartheta}_{i} & -ir\sin  {\vartheta}_{i} \\
0 & t & -ir\sin  {\vartheta}_{i} & r\cos  {\vartheta}_{i} \\
 \end{array} \right),
 \end{equation}
where $r$ and $t$ are  {respectively} the reflection and transmission amplitudes at the beam splitter. $\vartheta_{i}$ is a parameter that depends on the angle between the beam splitter and the leads (incidence angle) and the spin orbit parameters\cite{Lopez}. In reference  [{\onlinecite{Feve}}] the authors derived a similar $S$ matrix for a beam splitter in a Rashba active medium. In that case, very conveniently, the $S$ matrix is independent of the spin orbit parameters. In the more general case of a medium with both Rashba and Dresselhaus SO couplings, $ {\vartheta}_{i}$ becomes independent of the spin-orbit parameters and equal to the incidence angle  in the limit in which  it approaches $\frac{\pi}{4}$.\cite{Lopez} Reflection at the beam splitter mixes spin orientations due to the change in direction of the $\vec{k}$ vector within spin-orbit active media, changing the orientation of the effective wave-vector dependent magnetic field. Therefore, reflection matrices in the spin basis are non-diagonal.{\cite{Lopez}} In contrast, direct transmission (no change in direction of $\vec{k}$ vector) does not mix spin states, thus, transmission is represented by diagonal matrices. We will consider a $50-50$ beam splitter, with equal transmission and reflection probabilities and incidence angle fixed at $\frac{\pi}{4}$. In this case, the state after the collision with the beam splitter is
\begin{equation}
\vert \Psi \rangle =-\frac{1}{\sqrt{2}}\left( a_{3+}^{\dagger}a_{4+}^{\dagger}+a_{3-}^{\dagger}a_{4-}^{\dagger} \right)\vert 0 \rangle.
\end{equation}
In order to translate the beamsplitter output along the leads one must build the orbital and precessional operator from the general form\cite{Lopez}
\begin{equation}
\exp{\Big [\frac{i}{\hbar}\int_{{\mathcal L}}{d {\bf l}\cdot}({\bf p}-\frac{g\bfa^a\sigma^a}{2})\Big]},
\label{EvolutionOp}
\end{equation}
where $\bf p$ is the momentum, and one integrates along path $\cal L$, and
\begin{equation}
\frac{g}{m}\bfa^a\sigma^a
= (\beta\sigma^x-\alpha\sigma^y)\hat {\mathbf x}+(\alpha\sigma^x-\beta\sigma^y)\hat {\mathbf y},
\label{SU2VectorPotential}
\end{equation}
where $m$ is the effective mass of the electron in the electron gas. Recalling the definition of $\kappa=\arctan\frac{\beta}{\alpha}$, one can
identify $\alpha=\cos\kappa$ and $\beta=\sin\kappa$. Thus, the electron wave function is translated along lead 3 through a distance $L$, in the $y$ direction (picks out $\hat {\bf y}$ in Eq.\ref{SU2VectorPotential}), by operating on the state with the spin-dependent matrix as $|\Phi_3(L) \rangle=e^{-i\Lambda  {L} \sigma^{1}}a^{\dagger}_{3\pm}|0\rangle$, where
$\Lambda=\frac{ {m}\sqrt{\alpha^2+\beta^2}}{2\hbar}$  and $\sigma^{1}=\sigma^{x}\cos {\kappa} - \sigma^{y}\sin  {\kappa} $. The electron that travels through the lead 4 is spatially evolved with the matrix $e^{-i\Lambda L \sigma^{2}}$, where
$\sigma^{2}=\sigma^{x}\sin  {\kappa} -\sigma^{y}\cos {\kappa} $. Applying the translation matrices to the creation operators, we verify that $a_{3\pm}^{\dagger}$ ($a_{4\pm}^{\dagger}$) create eigenfunctions of the operator $e^{-i\Lambda {L} \sigma^{1}}$ ($e^{-i\Lambda {L} \sigma^{2}}$) with eigenvalues $e^{\pm i\Lambda L}$. The orbital part of the operator in Eq.\ref{EvolutionOp} only contributes a global phase that can be dropped.

On lead 4 we place a junction that symmetrically couples and electron reservoir to both components of the spin-orbit basis spinors. The reservoir injects electrons through a perfect wire described by the single electron wavefunction $\Psi (y)=\sqrt{N}(e^{iky}+C e^{-iky})$ where $N= e f(E)/2\pi \hbar v$  with the Fermi distribution {
$f(E)=\left[\exp ({E-E_{F}})/{kT}+1\right]^{-1}$}, $v=\hbar k/m$, and $C$ is the amplitude reflected back into the reservoir. There are three incoming  {amplitudes (denoted as $\cal A$, $\cal B$ and $\cal C$)} and three outgoing channels  {(${\cal A}'$, ${\cal B}'$ and ${\cal C}'$)} at the junction. The first amplitude refers to the reservoir wire, and the latter two other to lead 4. Thus the scattering matrix that describes the collision at the junction is a $3\times 3$ matrix \cite{Buttiker,Buttikeretal}:
\begin{equation} \left( \begin{array}{c}  {{\cal A}}^{\prime} \\  {{\cal B}}^{\prime} \\  {{\cal C}}^{\prime} \\ \end{array} \right)=\left( \begin{array}{ccc}
-(a+b) & \sqrt{\epsilon} & \sqrt{\epsilon}  \\
\sqrt{\epsilon} & a & b \\
\sqrt{\epsilon} & b & a \\
 \end{array} \right) \left( \begin{array}{c}  {{\cal A}}\\  {{\cal B}} \\ 0 \\ \end{array}\right),
 \end{equation}
 with $a=\frac{1}{2}\left(\sqrt{1-2\epsilon}-1\right)$ and $b=\frac{1}{2}\left(\sqrt{1-2\epsilon}+1\right)$.
 $\epsilon$ is a coupling parameter (the maximum coupling of the reservoir to the lead is achieved by the value $\epsilon=1/2$ and
 $\epsilon=0$ is the decoupled limit\cite{Buttiker}).

In the wave-function along the wire $\Psi (y)$ given above and which connects the device to the reservoir, the reservoir injects
$\sqrt{N}$ and gets back the amplitude $\sqrt{N}C$. The constant $C$ can be univocally determined from scattering matrix system of equations as
\begin{equation}
C= -(a+b)+\sqrt{\frac{\epsilon}{N}}\exp\left ({\frac{im\sqrt{\alpha^2+\beta^2+\frac{2E}{m}}}{\hbar}}\right ).
\end{equation}
The current along the {wire, defined by $j_d=\Psi^{\dagger} ev \Psi$,} is $j_d=N(1-\vert C\vert ^2)$. The zero current condition is satisfied if $\vert C\vert^2=1$ which would provide a pure dephasing source\cite{Buttiker}. In contrast to the single channel case\cite{Buttiker}, the spinor nature (up down spin-orbit basis) of the state in lead 4 permits implementing a net loss/gain of current moderated by the reservoir voltage and the coupling parameter $\epsilon$. We are then able to implement a lossy beamsplitter arm capable of modifying the information content of the beam splitter state. The lost/gained current satisfies the equation
\begin{equation}
\frac{1-2N}{2\sqrt{N}}+\frac{j_d}{2\sqrt{N}\epsilon}=\sqrt{\frac{1-2\epsilon}{\epsilon}}\cos\left(\frac{m}{\hbar}\sqrt{\alpha^2+\beta^2+\frac{2E}{m}}\right).
\label{lostcurrent}
\end{equation}
Notice that the parameter {$N$} on the left hand side of this equation depends on the characteristics of the reservoir (temperature, Fermi energy), whereas the parameters on the right hand side are related to the spin-orbit parameters of the device. Therefore, we can see that in order to avoid dissipation ($j_d=0$), a particular relation between all these parameters needs to be satisfied. We will implement the reservoir model at a particular energy  in the vicinity of the Fermi energy of the reservoir. All the scattering with the reservoir junction is computed at this energy.

The collisions of the electrons in arm $4$ with the reservoir junction scatter according to Eq.\ref{EvolutionOp} and \ref{SU2VectorPotential} as
\begin{eqnarray}
e^{i\Lambda L}a^{\dagger}_{4+}\vert 0 \rangle \rightarrow&&\frac{\Phi_{y}(x,y)}{\sqrt{2}}[ r_{+}e^{i\Lambda L}e^{ik_{+}L}b^{\dagger}_{4\uparrow}+ \nonumber \\
&&e^{i\Lambda L}(t_{+}a^{\dagger}_{4\uparrow}+ ie^{i\kappa}a^{\dagger}_{4\downarrow})]\vert 0 \rangle, \
\end{eqnarray}
\begin{eqnarray}
e^{-i\Lambda L}a^{\dagger}_{4-}\vert 0 \rangle \rightarrow&&\frac{\Phi_{y}(x,y)}{\sqrt{2}} [ r_{-}e^{-i\Lambda L}e^{ik_{-}L}b^{\dagger}_{4\uparrow}+\nonumber \\
&&e^{-i\Lambda L}(t_{-}a^{\dagger}_{4\uparrow}-ie^{i\kappa}a^{\dagger}_{4\downarrow}) ]\vert 0 \rangle. \
\end{eqnarray}
The outgoing state must be expressed as a function of the least possible number of operators in the spin-orbit basis, in order to simplify the computation of the entanglement and polarization, therefore we express the reflected operators in terms of the operator of electrons that travel in the positive direction of arm 4 via the relations $b^{\dagger}_{4+}=e^{-i(k_{+}+k_{-})x}a^{\dagger}_{4-}$ and $b^{\dagger}_{4-}=e^{-i(k_{+}+k_{-})x}a^{\dagger}_{4+}$. After performing this change, and transforming the expressions to the spin-orbit basis, the states of the electrons after collision with the reservoir junction may be written in the form
\begin{eqnarray}
e^{i\Lambda L}a_{4+}^{\dagger}\vert 0 \rangle&& \rightarrow \frac{1}{2}e^{i\Lambda L}\Big [(t_{+}+r_{+}+1)a_{4+}^{\dagger}+\nonumber \\
&&(t_{+}+r_{+}-1)e^{i(k_{+}-k_{-})L}a_{4-}^{\dagger} \Big ]\vert 0 \rangle, \\
e^{-i\Lambda L}a_{4-}^{\dagger}\vert 0 \rangle&&\rightarrow \frac{1}{2}e^{-i\Lambda L}\Big [(t_{-}+r_{-}-1)e^{-i(k_{+}-k_{-})L}a_{4+}^{\dagger}\nonumber \\
&&+(t_{-}+r_{-}+1)a_{4-}^{\dagger} \Big ]\vert 0 \rangle,
\end{eqnarray}
where
\begin{eqnarray}
r_{+}&=&\sqrt{N\epsilon}e^{-i\Lambda L}e^{-ik_{+}L}+a, \nonumber \\
r_{-}&=&\sqrt{N\epsilon}e^{i\Lambda L}e^{-ik_{-}L}+a,
\end{eqnarray}
are the electron reflection amplitudes, and
\begin{eqnarray}
t_{+}&=&\sqrt{N\epsilon}e^{-i\Lambda L}e^{-ik_{+}L}+b,\nonumber \\
t_{-}&=&\sqrt{N\epsilon}e^{i\Lambda L}e^{-ik_{-}L}+b,
\end{eqnarray}
are the transmission amplitudes.  The output normalized state is
\begin{equation}
\vert \Psi \rangle =\frac{\left[X a_{3+}^{\dagger}a_{4+}^{\dagger}+Y a_{3+}^{\dagger}a_{4-}^{\dagger}+Za_{3-}^{\dagger}a_{4+}^{\dagger}+Wa_{3-}^{\dagger}a_{4-}^{\dagger}\right]}{\sqrt{X^2+Y^2+Z^2+W^2}}\vert 0 \rangle,\\
\label{OutputStateRes}
\end{equation}
where
\begin{eqnarray}
X&=&-\frac{1}{2\sqrt{2}}e^{2i\Lambda L}(t_{+}+r_{+}+1),\nonumber \\
Y&=&-\frac{1}{2\sqrt{2}}(t_{+}+r_{+}-1),\nonumber \\
{Z}&=&-\frac{1}{2\sqrt{2}}(t_{-}+r_{-}-1),\nonumber \\
{W}&=&-\frac{1}{2\sqrt{2}}e^{-2i\Lambda L}(t_{-}+r_{-}+1).
\label{XYWZ}
\end{eqnarray}
Knowing the outgoing state we can easily compute the degree of entanglement of the output.
In order to compute the concurrence, a robust entanglement measure, we need to identify  the $\Omega$ matrix\cite{Lopez2} in the expansion $\vert \Psi \rangle=\sum_{l,m}
\Omega_{l,m}c^{\dagger}_{l}c^{\dagger}_{m} \vert 0\rangle$, where {$l,m\in \{1,2,3,4\}$}
with the correspondance $\{ c^{\dagger}_{1},c^{\dagger}_{2},c^{\dagger}_{3},c^{\dagger}_{4}\}\rightarrow \{
a^{\dagger}_{3+},a^{\dagger}_{3-},a^{\dagger}_{4+},a^{\dagger}_{4-}\}$. Writing the outgoing state as $\vert \Psi
\rangle=(\gamma_{11}c^{\dagger}_{1}c^{\dagger}_{3}+\gamma_{12}c^{\dagger}_{1}c^{\dagger}_{4}+\gamma_{21}c^{\dagger}_{2}c^{\dagger}_{3}+\gamma_{22}c^{\dagger}_{2}c^{\dagger}_{4})\vert
0 \rangle$, where $\gamma_{ij}$ can be readily identified from Eq.~\ref{OutputStateRes}, the $\Omega$ matrix is given by
\[ \Omega=\left( \begin{array}{cccc}
0 & 0 & \gamma_{11} & \gamma_{12}   \\
0 & 0 & \gamma_{21} & \gamma_{22}\\
0 & 0 & 0 & 0 \\
0 & 0 & 0 & 0 \\
 \end{array} \right).\]
We compute the skew-symmetric matrix $W^{S}=\frac{1}{2}(\Omega-\Omega^{T})$. The expression for the concurrence is
$\eta=\epsilon^{\alpha\beta\mu\nu}W^{S}_{\alpha\beta}W^{S}_{\mu\nu}$, where $\epsilon^{\alpha\beta\mu\nu}$ is the
totally antisymmetric unit tensor in four dimensions. Then $\eta_{\rm out}=8\vert
W^{S}_{12}W^{S}_{34}+W^{S}_{13}W^{S}_{42}+W^{S}_{14}W^{S}_{23} \vert=2\vert  \gamma_{12}\gamma_{21}-\gamma_{11}\gamma_{22} \vert$. The concurrence of the outgoing state is given by
\begin{equation}
 \eta_{out}=\frac{\left| t_{+}+r_{+}+t_{-}+r_{-} \right|}{2(\vert X\vert ^2+\vert Y\vert ^2+\vert Z \vert ^2+\vert W\vert^2)}.
\end{equation}
\begin{figure}
\includegraphics[width=8.3cm]{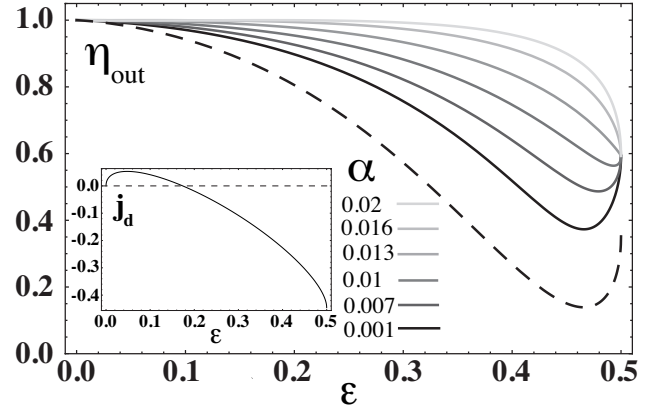}
\caption{\label{fig2} Plot of the concurrence (shades of gray) and linear entropy (dashed line) of the outgoing electrons as a function of coupling to the reservoir $\epsilon$  for the spin orbit strengths indicated ($\beta=0.004$ a.u). The linear entropy is evaluated for the smallest value of the SO interaction. The energy of the injected electrons is $E=0.2$ {a.u.} at the {Fermi} energy of the reservoir at a reference temperature of $~90$K. The inset shows the decoherence current as a function of $\epsilon$ from Eq.\ref{lostcurrent}.}
\end{figure}
\begin{figure}
\includegraphics[width=8cm]{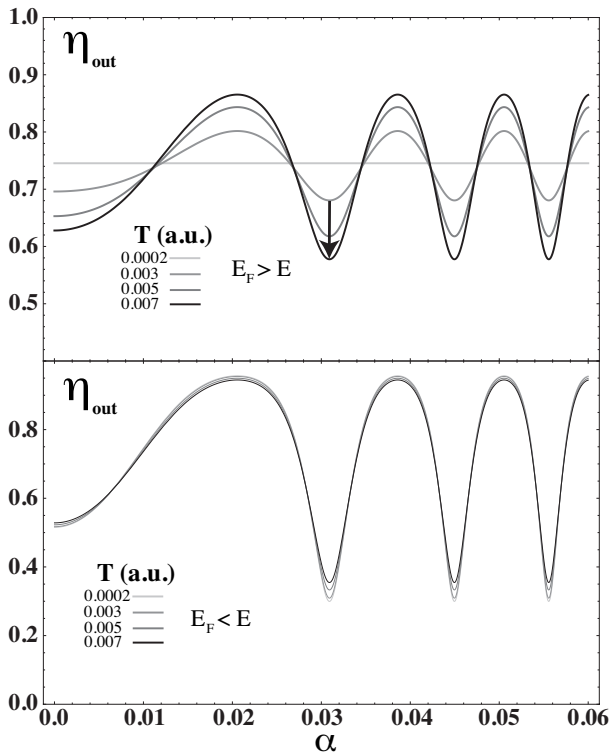}
\caption{\label{fig3} Concurrence versus the Rashba coupling parameter, for $\beta=0.004$ a.u., Fermi energy $E_F=0.21 a.u$ (top panel) $E_F=0.19$ a.u (bottom panel) while the electron energy in the interferometer arm is $E=0.2 a.u$. Depending on the electron energy, temperature can either enhance concurrence oscillations (top) or reduce them (bottom). The temperature range has been chosen rather large to enhance the visibility of the effect. Arrow indicates the direction of temperature increase. $T=3.157\times 10^5K=1$ a.u.}
\end{figure}
The concurrence measures the degree of entanglement with a scalar between zero (minimum) and one (maximum). It is monotone with the more intuitive measure of the subspace entropy $S_{L}$ that quantifies the purity of a state. We also compute the linear entropy of one of the particle subspaces to obtain an alternative measurement of entanglement. In order to obtain the linear entropy one must first compute the density matrix of the full outgoing state, then calculate the partial trace over the state of one of the electrons (e.g. labeled $a$) to obtain the density matrix $\rho_{T}={\rm tr}_{a}\rho_{\rm out}$ of the other electron that arrives at the detector. The linear entropy is then computed as $S_{L}=2(1-{\rm tr} \rho_{T}^2)$.

The concurrence and linear entropy, of the outgoing electrons for different values of spin-orbit coupling, are plotted as a function of the coupling to the electron reservoir in Fig.~\ref{fig2}. The particular values chosen for the spin-orbit parameters reflect order of magnitude values for typical semiconducting materials where such a device could be implemented ($3.9\times 10^{-12}$ eV.m
=0.0027 a.u.\cite{Takayanagi}). The concurrence, in the figure, rises from the dark to the lightly shaded curves as the SO increases and oscillates, in this range, with further changes in the coupling (See Fig.\ref{fig3} which also exhibits the temperature dependence). The value of the bulk Dresselhaus coupling is kept fixed at a physical value as it is generally not manipulated through the common experimental knobs applicable to 2D electron gases. Nevertheless, the behavior upon varying $\beta$ is symmetrical to the behavior as a function of $\alpha$ the Rashba parameter. So we will keep the Dresselhaus interaction fixed throughout the rest of this work.
Note that pure dephasing i.e. $j_d=0$ illustrated in the inset of Fig.\ref{fig2}, is not enough to generate changes in entropy of the beamsplitter state function. Magnification of the concurrence in the figure shows that it recovers the maximum value when $j_d$ crosses zero value.

The reason for the dependence of the concurrence with the SO coupling is due to the spinor components relative phase dependence on the SO strength and position. Such a dependence modulates the coupling to the reservoir through the wavefunction amplitude. We have shown this dependence explicitly by probing voltage at different points of the beam splitter arm, observing characteristic oscillations for a fixed value of SO strength. Thus both SO strength and reservoir position can be used to modulate entanglement.
\begin{figure}
\includegraphics[width=8.7cm]{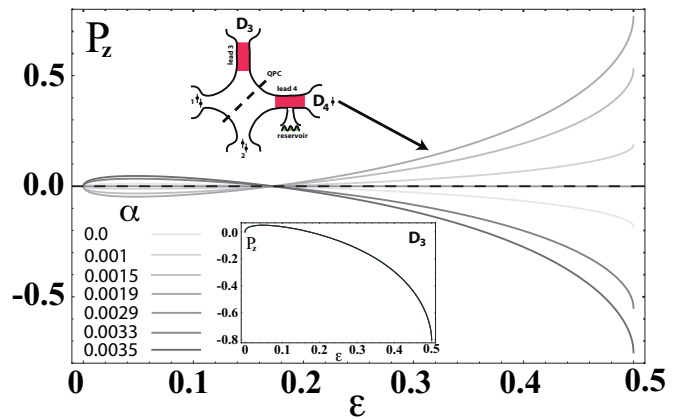}
\caption{\label{fig4} The $z$ component of the polarization for the outgoing electrons at detector  $D_{4}$ as a function of coupling to the reservoir $\varepsilon$ and for representative values of Rashba spin-orbit coupling. The output state is completely unpolarized for $\epsilon\rightarrow 0$. As current is leaked from or to the reservoir, a polarization arises that depends on the spin orbit strength of the SO active region. The Dresselhaus interaction is set to 0.004 a.u. The inset shows the $z-$component of the polarization at the other detector for $\alpha=0.0019$.}
\end{figure}

The minimum of the concurrence at a particular value of $\epsilon$, is an intriguing feature since the reservoir appears to loose its ability to reduce the entanglement of the beam splitter state. This reflects a rather non-ideal nature of the voltage probe model: The leads coupling to the reservoir carry parameter dependent entropy that can be computed explicitly, and there is an exchange of entropy between the arm mixed state and the leads to the reservoir. Such electron density in the reservoir lead, partly stores and cedes information to the two electron system (see this effect in another context in ref.\onlinecite{Lopez2}). This is consistent with the behavior of the persistent current states in contact with such a reservoir model e.g. it cannot yield complete decoherence at any finite temperature\cite{Buttiker}. In spite of this feature the current reservoir model allows for a proof of concept with a full analytical treatment.

\section{Spin-polarization for electrons in an entangled state}
Since the incoming electrons at the beam splitter are in a maximally entangled spin singlet state, the incoming polarization is zero. The electron on, say, port three of the pair is then in a completely mixed state $\rho_{\rm in}^{3}=\frac{1}{2}\left( \ket\uparrow \bra\uparrow+ \ket\downarrow \bra\downarrow\right)$ i.e. proportional to the identity matrix, e.g. has zero polarization and maximum entropy. For the entangled state considered, there is complete anti bunching so that there is one electron per arm as seen from Eq.\ref{OutputStateRes}.

Calculating the density matrix of the full outgoing state $\rho_{\rm out}=\vert \Psi_{\rm out}\rangle \langle \Psi_{\rm out}\vert$ and taking the partial trace over the electron on the other arm (port 4), to obtain the density matrix of the electron on port 3, we can compute the components of the polarization of the outgoing electron beam received at detector $D_{3}$:
\begin{eqnarray}
 {P^{x}_{\rm out}}&=&AC^{*}+CA^{*}+BD^{*}+DB^{*}, \\
 {P^{y}_{\rm out}}&=&i(AC^{*}-CA^{*}+BD^{*}-DB^{*}),\nonumber \\
 {P^{z}_{\rm out}}&=&\vert A \vert ^2+\vert B \vert ^2-\vert C \vert ^2-\vert D \vert ^2\nonumber,
\end{eqnarray}
where
\begin{eqnarray}
A&=&\left( X+Y+Z+W \right)/2,
 \\
B&=&ie^{i\kappa}\left( X-Y+Z-W \right)/2,
 \nonumber \\
C&=&e^{-i\kappa}\left( -X-Y+Z+W \right)/2,
 \nonumber\\
D&=&i\left( -X+Y+Z-W \right)/2\nonumber.
\end{eqnarray}
$X,Y,W,Z$ were defined in Eq.~\ref{XYWZ}.  A plot of the $z-$component of the polarization of the electrons at detector $D_{4}$ as a function of coupling to the electron reservoir and for different SO couplings is shown in Fig.~{\ref{fig4}}. The figure shows how the change in entropy of the one electron subspace, introduced by the reservoir, changes the average polarization at the detector. It also displays how one can manipulate the polarization between positive and negative values by changing the Rashba SO coupling at the arm. It is important to note that while the concurrence is computed with the full wavefunction, including the reflected component from the electron reservoir, the polarization is only computed from the electron arriving at $D_4$. Thus a post-selection is implied by ignoring the reflected probability current, and this of course contributes to the resulting polarization. Note the perfect correlation or anti correlation to the decoherent current of the polarization curves depending on the Rashba parameter and the coupling to the reservoir, including the zero current situation where no polarization is observed.

The corresponding polarization at detector $D_3$ is shown in the inset of Fig.\ref{fig4}, the polarizations in the $x$ and $y$ components are negligible for this arm for the range of SO considered. This is due to particular choice of coupling to the reservoir at lead 4 and the non locality of the entangled state. While the coupling is symmetric in the SO basis, when translated into the spin basis, the same coupling results only to one component of spin (mixed by the SO interaction). This certainly is a source of bias toward the effects of a particular polarization. The polarization in detector $D_4$ nevertheless exhibits all components close in magnitude. Nevertheless, the magnitude of the polarization is a measure of entanglement itself so it should be the same for both detectors. This is verified by our expressions. The degree of freedom to polarize in different spin axes is then dependent on the details of the connection to the reservoir, and this could be exploited further as a control device.
\begin{figure}
\includegraphics[width=8.5cm]{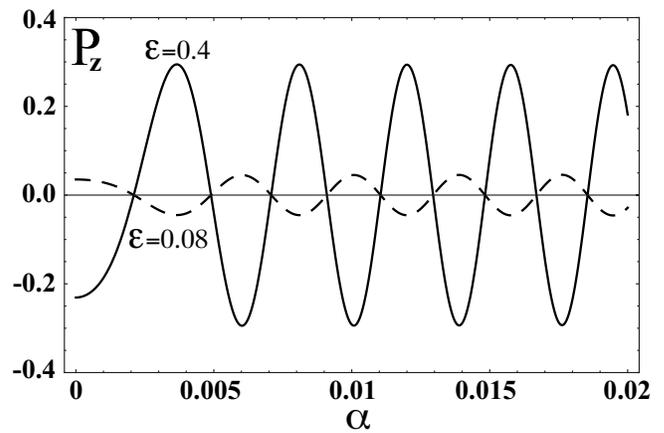}
\caption{\label{fig5} The $z$ component of the polarization for the detector $D_{4}$ as a function of the Rashba SO parameter at two different couplings to the reservoir indicated in the figure. When current is injected to the beam splitter arm (before nodal point in Fig.\ref{fig3}, also inset in Fig.\ref{fig2}) the oscillation is $\pi$ out of phase compared to the situation when current is drawn.}
\end{figure}

Figure \ref{fig5} shows the oscillations in the $z-$component of the polarization in detector $D_4$ as a function of the Rashba parameter, for two different values of the coupling to the reservoir. While the magnitude of the polarization reflects the reservoir coupling, the phase of the oscillation reflects the sign of the decoherent current. This can be seen by comparing to the inset of Fig.\ref{fig2}. 

On the basis of the spin-orbit strengths involved in Fig.\ref{fig5} we can estimate the consistency of the single subband approach 
in accordance with the criterion in ref.\cite{Egues}. In terms of our description, the criterion is that $\alpha |\langle \Phi_a|\partial_y|\Phi_b\rangle| << \varepsilon_b-\varepsilon_a$ where  $\Phi_{a,b}$ are the transverse confined wave function defined in section II, $\alpha$ the Rashba coefficient and $\varepsilon_{a,b}$ the corresponding subband energies. For a hard wall potential the matrix element in the previous expression can be easily evaluated  to $|\langle \Phi_a|\partial_y|\Phi_b\rangle|=8/3w$ where $w$ is the width of the beamsplitter leads. Using typical values of the Rashba coefficient in Fig.\ref{fig5} ($\alpha=0.003$ a.u) one arrives at $w< 1850$ a.u. equivalent to $\sim 100$ nm. This length scale is realizable with current lithography and gating techniques.

\section{Spin-polarization of electrons in a completely mixed state}
We will now consider that the incoming state of the electrons is prepared as a perfect mixture given by the density matrix
\begin{equation}
\rho_{\rm in}=\frac{1}{2}\ket{\ \!1+} \bra{1+}+\frac{1}{2}\ket{\ \!1-}\bra{1-},
\label{completemix}
\end{equation}
where $\vert 1+ \rangle=a_{1+}^{\dagger}\vert 0 \rangle$ and $\vert 1- \rangle=a_{1-}^{\dagger}\vert 0 \rangle$.
The density matrix is $\rho_{\rm in}=\frac{1}{2}\mathds{1}$, proportional to the identity so that the incoming polarization is zero. The electron is incoming from arm 1 of the beamsplitter and is scattered into both outgoing arms. There are no non-local correlations between the
output arms of the beamsplitter in this case.

After collision with the beam splitter, the states
$ {a_{1+}^{\dagger}}\vert 0 \rangle$ and $ {a_{1-}^{\dagger}}\vert 0
\rangle$ become
\begin{eqnarray}
\vert 1+ \rangle&\rightarrow& \frac{1}{\sqrt{2}}\left(
\frac{i}{\sqrt{2}}a_{3+}^{\dagger}-\frac{1}{\sqrt{2}}a_{3-}^{\dagger}+a_{4+}^{\dagger} \right)\vert 0 \rangle,  \\
\vert 1- \rangle&\rightarrow& \frac{1}{\sqrt{2}}\left(
-\frac{1}{\sqrt{2}}a_{3+}^{\dagger}+\frac{i}{\sqrt{2}}a_{3-}^{\dagger}+a_{4-}^{\dagger} \right)\vert 0 \rangle.
\end{eqnarray}
After including propagation through leads $3$ and $4$ and the coupling to the electron reservoir, according to the
discussion of the previous section, one has
 \begin{widetext}
\begin{eqnarray}
\label{Eq1+out}\vert 1+,{\rm out}\rangle&=& \frac{1}{\sqrt{2}}\left(
\frac{i}{\sqrt{2}}e^{i\Lambda L}a_{3+}^{\dagger}-\frac{1}{\sqrt{2}}e^{-i\Lambda L}a_{3-}^{\dagger}\right)\vert 0
\rangle
+\frac{1}{\sqrt{2}}\left(Xa_{4+}^{\dagger}+Ya_{4-}^{\dagger} \right)\vert 0 \rangle,
\\
\label{Eq1-out}\vert 1-,{\rm out}\rangle&=& \frac{1}{\sqrt{2}}\left(
-\frac{1}{\sqrt{2}}e^{i\Lambda L}a_{3+}^{\dagger}+\frac{i}{\sqrt{2}}e^{-i\Lambda L}a_{3-}^{\dagger}\right)\vert 0
\rangle 
+\frac{1}{\sqrt{2}}\left(Za_{4+}^{\dagger}+Wa_{4-}^{\dagger} \right)\vert 0 \rangle.
\end{eqnarray}
\end{widetext}
The density matrix of the outgoing state is then
\begin{equation}\rho_{\rm out}=\frac{1}{2}\vert 1+,{\rm out} \rangle \langle
1+,{\rm out}\vert+\frac{1}{2}\vert 1-,{\rm out} \rangle \langle 1-,{\rm out}\vert .\end{equation}
From the density matrix of the outgoing state in the spin basis, one obtains the outgoing polarization
$P^{z}={\rm tr}(\rho_{out}\sigma^{z})$. There is a new issue concerning the locality of the state in this case
which relates to the fact that wave functions in different output leads do not overlap. Thus, the cross terms that mix probabilities of traveling simultaneously through leads $3$ and $4$ do not contribute to the polarization. Let us consider for example the term $\frac{i}{\sqrt{2}}X^{*}e^{i\Lambda L}\vert 3+\rangle \langle 4+ \vert$, whose contribution to the full outgoing density matrix we will label $\rho_{D}$.
 {The states $\vert 3+ \rangle$ and $\vert 4+ \rangle$,}
\begin{eqnarray}
\vert 3+ \rangle&=& \frac{\Phi_{x}(x,y)}{\sqrt{2}}(\vert \uparrow\rangle -e^{-i {\kappa}}\vert \downarrow\rangle),
  \\
\vert 4+ \rangle&=& \frac{\Phi_{y}(x,y)}{\sqrt{2}}(\vert \uparrow\rangle -ie^{i {\kappa}}\vert \downarrow\rangle),
\end{eqnarray}
 allow one to write $\rho_D$ in the spin basis as
\begin{equation}
\rho_{D}=\frac{iX^{*}e^{i\Lambda L}\Phi_{x}(x,y)\Phi_{y}(x,y)}{2\sqrt{2}}e^{ikx}e^{iky}
\left(\begin{array}{cc} 1 & -ie^{-i {\kappa}} \\ -e^{-i {\kappa}} &  ie^{-i2 {\kappa}} \end{array}\right).
\end{equation}
In order to compute the contribution of $\rho_{D}$ to the $z$ component of the polarization we must {evaluate} the trace over the internal spin space and the coordinate space to yield
\begin{equation}
P_{D}^{z}=\frac{iX^{*}e^{i\Lambda L}}{2\sqrt{2}}(1-ie^{-2i\kappa})\int dx dy \Phi_{x}(x,y)\Phi_{y}(x,y)e^{ik(y-x)}.
\end{equation}
The integration of the product of the confining wave functions  vanishes since the wave function on the leads do not overlap. Therefore, the terms in the density matrix that combine electrons traveling through different leads do not contribute to the spin polarization. The final form of the output density matrix contains terms representing electrons that are collected at detectors $D_{3}$ and $D_{4}$. Since physically, polarization will only be measured either at one detector or the other, we must treat the density matrix at each detector using the above prescription. The density matrix of the outgoing state at detector $D_{3}$ is then
\begin{figure}
\includegraphics[width=8.5cm]{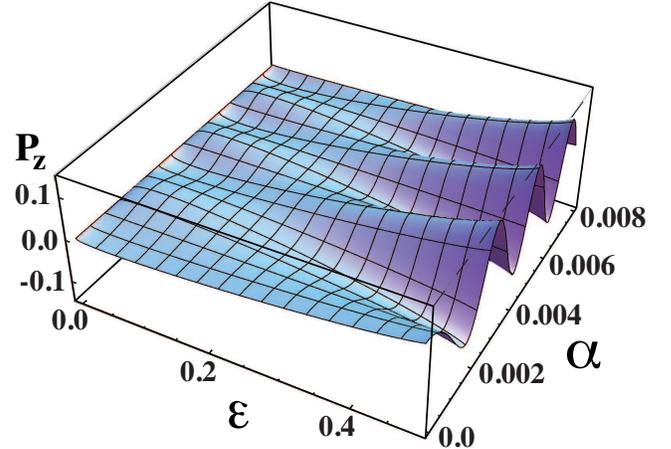}
\caption{\label{fig6}  The polarization of the outgoing state as a function of coupling to the electron reservoir $\epsilon$ for different values of the Rashba parameter. The energy is fixed at the Fermi energy of the reservoir and the temperature of the reservoir is $90$K. The position of the reservoir junction is set to $x=1\mu$m (see Fig. 1)}
\end{figure}
\begin{eqnarray}
\rho_{\rm out}^{D_{3}}=\frac{1}{2}\left( \vert 3+ \rangle \langle 3+ \vert + \vert 3- \rangle \langle 3- \vert \right)
\end{eqnarray}
This is the density matrix of a completely mixed state, therefore the polarization of the electron beam collected at detector $D_{3}$ is zero. This is expected from the lack of correlation between the leads. No action of the reservoir should be seen at detector $D_{3}$. Meanwhile, at detector $D_{4}$
\begin{eqnarray}
&&\rho_{\rm out}^{D_{4}}=\nonumber \\
&&\frac{1}{4}\Bigg ( (\vert X\vert^{2}+\vert Z\vert^2)\vert 4+\rangle \langle4+\vert + (\vert Y\vert^2 +\vert W\vert^2)\vert 4- \rangle \langle 4-\! \vert \nonumber \\
&&+ (XY^* + ZW^*) \vert 4+ \rangle \langle 4-\! \vert
+ (X^{*}Y+WZ^{*})\vert 4-\rangle \langle 4+\!\vert\Bigg ).\nonumber \\
\end{eqnarray}
Using Eq.\ref{XYWZ} we can develop the full expression in terms of SO and reservoir parameters. It is clear from the expression that we can tune the entropy of the mixed state on this lead by manipulating the state vector coefficients. While the analytical expressions for the polarization are rather cumbersome to write out here, we show the results for the z component of the Polarization in Fig.~\ref{fig6}. The figure shows that one can generate polarization from a mixed state as a function of the reservoir coupling, and modulate it to positive or negative values by adjusting the SO interaction in a reasonable parameter range. As with the entangled state, the polarization depends on a finite decoherence current. Pure dephasing will not generate polarization. Here polarizations are smaller than those of the entangled state, because half of the electron amplitude is unaffected by the reservoir and remains unpolarized. This aside from the fact that only one electron is operating in this latter setup.

It is important to reemphasize here, that while a time reversal symmetric situation, the SO interaction will only redistribute polarization, so as to maintain the global polarization equal to that of the initial state, here the reservoir breaks time reversal symmetry. The lack of time reversal symmetry allows for changing the global polarization. This is the same mechanism implied when considering more than one sub-band in the leads\cite{Nikolic2}. There, the decoherence mechanism is the irreversible loss of information of the spin subspace due to coupling to am "orbital reservoir" of two sub-bands. The polarization can thus be degraded.  

\section{Summary}

We have proposed a device to generate spin-polarized currents on the basis of entropy changes to the incoming mixed and entangled states. The mechanism involved is quite different from information preserving schemes, where the total input spin is constant and spin components are separated in either a ``Stern-Gerlach" approach or in a controlled precession mixing SO and magnetic fields. Our proposal is based on the decoherence effects produced by  the connection of an electron reservoir to one of the spin-orbit active leads in a beam splitter arrangement. Polarization of the incoming state can be changed due to breaking of time reversal symmetry introduced by the reservoir.  When the incoming electrons are in an entangled state, the beam splitter device produces spin-polarized currents at both detectors. The magnitudes of the polarization vectors at the two detectors are identical but, in general, the components of the vector can be tuned. The global change in polarization is itself a measure of entanglement degradation. In the case of incoming electrons in a completely mixed state, the same beam-splitter device produces spin-polarized currents only at the detector located on the lead connected to the electron reservoir, whereas  the other detector  collects an unpolarized current. The change in SO parameters in both case produce polarization oscillations because spin precession generates a variable amplitude at the reservoir junction. The frequency of the oscillations depends on the spin-orbit parameters of the material and the position of the reservoir junction.

\acknowledgments
This work was supported by CNRS-Fonacit grant PI-2008000272, and by the support of Nancy-Universit\'e through an Invited Professor position (EM). LGA and EM acknowledge illuminating discussion with Rodrigo Medina.





\end{document}